\documentclass[12pt]{article}
\usepackage{amsmath,amssymb,epsfig}
\usepackage{graphicx,floatflt,subfigure}
\usepackage{epstopdf}
\usepackage{cite}

\makeatletter

\renewcommand\section{\@startsection {section}{1}{\z@}%
                                   {-3.5ex \@plus -1ex \@minus -.2ex}%
                                   {2.3ex \@plus.2ex}%
                                   {\normalfont\large\bfseries}}

\renewcommand\subsection{\@startsection{subsection}{2}{\z@}%
                                     {-3.25ex\@plus -1ex \@minus -.2ex}%
                                     {1.5ex \@plus .2ex}%
                                     {\normalfont\normalsize\bfseries}}

\newcommand{\im}{\text{Im}\,}
\newcommand{\re}{\text{Re}\,}
\newcount\hour \newcount\minute
\hour=\time \divide \hour by 60
\minute=\time
\def\now{%
\ifnum \hour<13
  \ifnum \hour=0 \advance \hour by 12 \number\hour:\else \number\hour:\fi%
     \ifnum \minute<10 0\fi%
     \number\minute%
\ A.M.%
\else \advance \hour by -12 \number\hour:%
  \ifnum \minute<10 0\fi%
  \number\minute%
  \ P.M.%
\fi%
}

\makeatother

\begin{document}

\baselineskip=18pt  
\numberwithin{equation}{section}  
\allowdisplaybreaks  



%
%


\thispagestyle{empty}

\vspace*{-2cm}
\begin{flushright}
{\tt arXiv:0712.3305}\\
CALT-68-2665\\
IPMU-07-0003
\end{flushright}

\vspace*{1.7cm}
\begin{center}
 {\large {\bf Metastable Vacua in Perturbed Seiberg-Witten Theories,}\\
{\bf Part 2:  Fayet-Iliopoulos Terms and K\"ahler Normal Coordinates}}

 \vspace*{2cm}
 Joseph Marsano,$^*$ Hirosi Ooguri,$^{*,\dagger}$
 Yutaka Ookouchi,$^*$ and Chang-Soon Park$^*$\\
 \vspace*{1.0cm}
 $^*$
{\it California Institute of Technology 452-48, Pasadena, CA 91125, USA}\\
~~\\
 $^\dagger$ {\it Institute for the Physics and Mathematics of the Universe, \\
University of Tokyo, Kashiwa 277-8582, Japan}\\[1ex]
 \vspace*{0.8cm}
\end{center}
\vspace*{.5cm}

\noindent
We show that the perturbation of an ${\cal N}=2$ supersymmetric
gauge theory by a superpotential linear in the K\"ahler normal coordinates
of the Coulomb branch, discussed in arXiv:0704.3613, is equivalent
to the perturbation by Fayet-Iliopoulos terms. It follows
 that the would-be meta-stable vacuum at the origin of
the normal coordinates in fact preserves ${\cal N}=1$ supersymmetry unless
the superpotential is truncated to a finite-degree polynomial of
the adjoint scalar fields.
We examine the criteria for supersymmetry breaking under a perturbation by Fayet-Iliopoulos terms and present a general classification of non-supersymmetric critical points.
In some explicit examples, we are also able to study local stability of these points and demonstrate that, if the perturbation is chosen appropriately, they indeed correspond to supersymmetry-breaking vacua.
Relations of these constructions to flux compactifications and geometric meta-stability are also discussed.

\newpage
\setcounter{page}{1} 





\section{Introduction}

The recent discovery of metastable vacua in supersymmetric QCD \cite{ISS}
has led to renewed interest in the subject of supersymmetry breaking.
Since then, metastable vacua have been found in a variety of supersymmetric theories,
and it has become increasingly clear that they are ubiquitous
in the space of supersymmetric field theories
(See \cite{ISreview} and reference therein). It has also
been realized that theoretical constraints on supersymmetry
breaking mechanisms can be circumvented if one accepts metastability
and that it gives a greater flexibility in model building
\cite{Kitano,DM,St4,CST,Murayama,Abel,CFS}.
 Furthermore, a number of these vacua have been successfully realized in
string theory \cite{OO,Kawano,Amariti,Ooguri:2006bg,St1,St2,St3,St4,St5,St6,St7,St8,St9},
where both the breaking of supersymmetry and the achievement of stability are described
geometrically.

One particularly simple way to engineer metastable vacua is to start with ${\cal{N}}=2$ supersymmetric gauge theory and deform it by a suitable superpotential \cite{Ooguri:2007iu,Pastras:2007qr}\footnote{See also \cite{Arai,Mazz} for meta-stable supersymmetry breaking vacua in Seiberg-Witten theories.}.  It has long been known that one can easily create non-supersymmetric critical points of the effective potential in this manner but stabilizing them requires more care. A key observation of \cite{Ooguri:2007iu} is that if one constructs the superpotential from a linear combination of K\"ahler normal coordinates \cite{AlvarezGaume:1981hn} associated to a point $u_0$ of the Coulomb branch, then not only does $u_0$ become a critical point but stability is guaranteed in a generic case
{\footnote{At nongeneric points there may be a flat direction (see \cite{Ooguri:2007iu}).}}.  In general, however, K\"ahler normal coordinates are given by an infinite series expansion in deviations away from $u_0$ \cite{Higashijima:2000wz} and they are not globally defined in the moduli space. In \cite{Ooguri:2007iu}, it was shown that one can truncate the series expansion to define a superpotential expressed as a finite-degree polynomial of the adjoint scalar fields in the ${\cal N}=2$ vector multiplet, which is globally defined and can still achieve stability.

Such a truncation may not seem to impact the physics to a great degree.  Nevertheless, we will demonstrate in this note that if one deforms the theory by a superpotential built from \emph{exact} K\"ahler normal coordinates then the supersymmetry-breaking vacuum at $u_0$ becomes instead a supersymmetry-preserving one.
On the other hand, if the superpotential is truncated to a finite-degree polynomial, the supersymmetry is genuinely broken at $u_0$; the particle spectrum at $u_0$ is still supersymmetric but interactions break supersymmetry. 
In this case, supersymmetry can be restored at a point where a massless dyon appears and the Coulomb branch metric becomes singular.

We will show that the superpotential given as a linear combination of
exact K\"ahler normal coordinates is identical to a specific combination of
electric and magnetic Fayet-Iliopoulos (FI)
terms, $a^i$ and $a_{D\,j}$,  of the low energy Abelian gauge
theory.  As such, what we land on is actually a classic model of
\emph{partial} supersymmetry breaking \cite{Antoniadis:1995vb,Ferrara}. This observation gives a fresh perspective on the supersymmetry-breaking vacua of \cite{Ooguri:2007iu} along with a natural understanding of their stability.  It also demonstrates quite explicitly why the K\"ahler normal coordinates fail to be globally well-defined.  They simply inherit the nontrivial monodromies of $a^i$ and $a_{D\,j}$ as one encircles singular points of the moduli space.

The fact that the vacuum at $u_0$ turns out to be supersymmetric does not preclude the existence of non-supersymmetric vacua elsewhere on the moduli space.  The full vacuum structure of theories with electric and magnetic FI terms has not
been completely understood, though, so one must embark on a more detailed analysis to establish the existence of metastable supersymmetry-breaking vacua in these models.

In the second part of this paper, we take some initial steps toward doing precisely this by
establishing a formalism for engineering nontrivial critical points in the perturbative regime of ${\cal{N}}=2$ theories deformed by electric and magnetic FI terms. In the simple example of a rank two gauge group, we are also able to directly address the problem of stability and demonstrate that stable non-supersymmetric vacua can be engineered in suitable parts of the perturbative regime.

While our focus in this paper is on deformed Seiberg-Witten theory, the structure considered makes a natural appearance in flux compactifications of type II superstring theory \cite{TaylorVafa,Vafa:2000wi,DouglasKachru}. In fact, it appears as a geometric engineering limit of such a compactification, where the coefficients $(e_i, m^i)$ of FI terms $W = \sum_i e_i a^i + m^i a_{D i}$ are identified with the amounts of fluxes. The scalar potential constructed from the superpotential $W$ is invariant under the monodromy transformation of $(a^i, a_{D i})$ provided the fluxes $(e_i, m^i)$ are also transformed appropriately. Thus, the potential is single-valued if we consider it as a function on the space of fluxes as well as on the Calabi-Yau moduli space. By contrast, in the field theory limit, the fluxes are frozen and become non-dynamical parameters, and the potential is multivalued in the Coulomb branch moduli space. This is caused since the field theory limit defined at a generic point in the Coulomb branch breaks down at massless dyon points because of the appearance of extra light particles at these points. It is exactly around each of these singular points where $W$ is multivalued in the field theory limit.

Because of this connection, the results in this paper can be used to classify non-supersymmetric critical points and study their stability in flux compactifications on local Calabi-Yau manifolds. In particular, there is a connection with recent studies of geometrically-induced stringy metastable vacua \cite{Aganagic:2006ex,Heckman:2007wk,Douglas:2007tu,Marsano:2007fe,Heckman:2007ub,Aganagic:2007kd,Aganagic:2007zm}.  Indeed, it was our interest in the system of \cite{Aganagic:2006ex} and its potential relationship to the vacua of \cite{Ooguri:2007iu} that formed the primary motivation for this work at the outset. A new feature of the class of models discussed in this paper is that supersymmetry breaking is taking place entirely in the field theory context. 

The organization of this note is as follows.  In section 2, we will demonstrate the connection between exact K\"ahler normal coordinates and FI parameters.  In section 3, we briefly review the structure of supersymmetry-preserving vacua in models with FI terms and their relation to the vacua of \cite{Ooguri:2007iu}.  In section 4, we consider the problem of engineering supersymmetry-breaking vacua in these models and explicitly demonstrate that this can indeed be done in the perturbative regime for the simple example of a gauge group with rank two.  In section 5, we comment on the relation to flux compactifications and recent work on supersymmetry-breaking in that context.  Appendix A contains some ${\cal{N}}=2$ superspace conventions and reviews the manner in which ${\cal{N}}=2$ supersymmetry is realized in the theories under study.

\section{K\"ahler Normal Coordinates and Fayet-Iliopoulos Terms}\label{sec:KNC}

In this paper, we shall devote our attention to generic points along the Coulomb branch of $SU(N)$ Seiberg-Witten theory \cite{SeibergWitten,KlemmLerche} where the IR physics is described by ${\cal{N}}=2$ $U(1)^{N-1}$ supersymmetric gauge theory with Seiberg-Witten prepotential ${\cal{F}}$. The Lagrangian of this theory can be written in ${\cal{N}}=1$ superspace as
\begin{equation}{\cal{L}}=\frac{1}{2}\im\left[\int\,d^4\theta\,{\cal{F}}_i(A_k)\bar{A}^i+\frac{1}{2}\int\,d^2\theta\,{\cal{F}}_{ij}(A_k)W_{\alpha}^iW^{\alpha\,j}\right]\label{modspaceact}\end{equation}
where $i=1\ldots N-1$ and ${\cal{F}}_{i_1i_2\ldots}=\partial_{i_1}\partial_{i_2}\ldots {\cal{F}}(a_i)$.  As usual, we often denote ${\cal{F}}_{ij}$ by $\tau_{ij}$, the period matrix of the Seiberg-Witten curve, and construct from this a K\"ahler metric on the Coulomb branch
\begin{equation}g_{i\bar{j}}=\im\tau_{ij}.\end{equation}

To this theory, let us consider adding a superpotential of the form
\begin{equation}W=k_iz^i,\label{KNCsup}\end{equation}
where the $z^i$ are a set of K\"ahler normal coordinates associated to a fixed point $a^i_0$.  As demonstrated in \cite{Ooguri:2007iu}, a study of the scalar potential of this theory in the vicinity of $a^i_0$ reveals that $a^i_0$ is (almost always) a stable critical point where the manifest ${\cal{N}}=1$ supersymmetry of the Lagrangian \eqref{modspaceact} plus superpotential is broken.

In general, the $z^i$ can be written in terms of special coordinates $a^i$ along the moduli space as \cite{AlvarezGaume:1981hn,Higashijima:2000wz}
\begin{equation}z^i = \Delta a^i + g^{i\bar{j}}(a_0)\sum_{n=2}^{\infty}\frac{1}{n!}\partial_{i_3}\ldots \partial_{i_n}\Gamma_{\bar{j}i_1i_2}(a_0) \Delta a^{i_1}\Delta a^{i_2}\ldots \Delta a^{i_n}\label{KNCdef}\end{equation}
where
\begin{equation}\Delta a^i \equiv a^i-a^i_0.\end{equation}
Because only the first few terms of \eqref{KNCdef} are needed to establish stability at $a^i_0$, one can follow \cite{Ooguri:2007iu} and truncate the series when constructing the superpotential \eqref{KNCsup}.  In this manner, it is possible to engineer long-lived metastable supersymmetry-breaking vacua by introducing an appropriate polynomial superpotential of finite degree.

In special coordinates, the connections $\Gamma_{\bar{j}i_1i_2}$ take a particularly
simple form
\begin{equation}\Gamma_{\bar{j}i_1i_2}=\frac{1}{2i}{\cal{F}}_{ji_1i_2}
=\frac{1}{2i}\partial_{i_2}\tau_{ji_1} = \frac{1}{2i} \partial_{i_1} \partial_{i_2}
a_{D\, j},\end{equation}
where $a_{D\, i} = \partial_i {\cal F}$.
This allows us to recognize the infinite series
in \eqref{KNCdef} as a Taylor expansion of $a_{D\,j}$ about
the point $a^i_0$.  In fact, we can easily
sum the series and write the exact K\"ahler normal coordinates $z^i$ as
\begin{eqnarray}
z^i &=& \Delta a^i + \left(\frac{1}{\tau_0 - \bar \tau_0}\right)^{ij}
      \sum_{n=2}^\infty \frac{1}{n!}
\frac{\partial^n a_{D\, j}(a_0)}{\partial a^{i_1} \cdots \partial a^{i_n}}
\Delta a^{i_1}\Delta a^{i_2}\ldots \Delta a^{i_n} \nonumber \\
&=& \Delta a^i + \left(\frac{1}{\tau_0 - \bar \tau_0}\right)^{ij}
\left(a_{D\, j}(a) - a_{D\, j}(a_0) - \tau_{0\,jk} \Delta a^k\right)
\nonumber \\
&=& \left(\frac{1}{\tau_0 - \bar \tau_0}\right)^{ij}
\left( a_{D\, j}(a) - \bar\tau_{0\, jk} a^k \right)
+ {\rm const},
\end{eqnarray}
where $\tau_{0\, ij} = \tau_{ij}(a_0)$.
This means that, up to irrelevant constant terms that we shall hereafter drop, the superpotential \eqref{KNCsup} is a specific linear combination of electric and magnetic FI terms
\begin{equation}W=e_ia^i+m^ia_{D\,i},\label{FIsup}\end{equation}
where
\begin{equation}
e_i = -k_j \left(\frac{1}{\tau_0-\bar{\tau}_0}\right)^{jk}\bar{\tau}_{0\,ki},
~\quad m^i =k_j \left(\frac{1}{\tau_0-\bar{\tau}_0}\right)^{ji}
\end{equation}
In particular the FI parameters satisfy
\begin{equation}e_i+m^j\bar{\tau}_{0\, ij}=0\end{equation}
We are therefore able to identify the theory with superpotential \eqref{KNCsup} as the classic model of partial supersymmetry breaking first introduced by Antoniadis et al \cite{Antoniadis:1995vb}\footnote{For local supersymmtric theories, see \cite{Ferrara}. }.  As we shall now review, the vacua at $a^i_0$ actually preserve an ${\cal{N}}=1$ supersymmetry, providing a natural explanation for their stability.

\section{Review of Supersymmetry-Preserving Vacua of ${\cal{N}}=2$ Abelian Theory with FI Terms}

In this section, we review the ${\cal{N}}=2$ formalism of the Abelian gauge theory \eqref{modspaceact} with superpotential \eqref{FIsup} and the conditions for having supersymmetry-preserving vacua.  This will allow us to see explicitly that the theory with superpotential \eqref{KNCsup} preserves the full ${\cal{N}}=2$ supersymmetry in an appropriate sense.  It will also make clear that the vacua at $a^i_0$ break the ${\cal{N}}=1$ supersymmetry  which is manifest in the superpotential formalism while preserving the ``hidden'' half.  The results of this section are well-known and included only for completeness{\footnote{We use the notation of  Antoniadis et al \cite{Antoniadis:1995vb}.  See \cite{Ivanov:1997mt} for an $SU(2)$-covariant approach that is equivalent.  This formalism has also been recently reviewed in \cite{Aganagic:2006ex}.}}.

\subsection{FI Terms and ${\cal{N}}=2$ Supersymmetry}

To clearly discuss how the non-manifest supercharges act, let us rewrite the action using ${\cal{N}}=2$ superspace.  For this, we introduce a second set of superspace coordinates $\tilde{\theta}$ and consider two types of ${\cal{N}}=2$ chiral superfields: a generic ${\cal{N}}=2$ chiral superfield ${\cal{A}}$ and a ``reduced'' ${\cal{N}}=2$ chiral superfield ${\cal{A}}_D$ which satisfies the constraint
\begin{equation}\left(D^{a\,\alpha}D^b_{\alpha}\right){\cal{A}}_D=\left(\overline{D}^a_{\dot{\alpha}}\overline{D}^{b\,\dot{\alpha}}\right){\cal{A}}_D^{\dag}\label{redconst}\end{equation}
This constraint ensures that the ``reduced'' superfield ${\cal{A}}_D$ contains only the component fields of the off-shell ${\cal{N}}=1$ chiral and vector multiplets.  The superfield ${\cal{A}}$, on the other hand, is unconstrained and hence contains several additional auxiliary fields.  For example, if we denote the $\theta^2$ component of ${\cal{A}}_D$ by $F_D$, the $\tilde{\theta}^2$ component is simply the complex conjugate $\bar{F}_D$.  By contrast, the $\tilde{\theta}^2$ component of ${\cal{A}}$, which we denote by $\tilde{F}$, has no \emph{a priori} relation to the $\theta^2$ component, $F$.  A more detailed review of our ${\cal{N}}=2$ superspace and superfield conventions and notation can be found in Appendix A.

To write FI terms, we introduce vectors of auxiliary components for both ${\cal{A}}$ and ${\cal{A}}_D$
\begin{equation}Y = \begin{pmatrix}i\left(F-\tilde{F}\right) \\ F+\tilde{F} \\ \sqrt{2}D\end{pmatrix}\qquad Y_D = \begin{pmatrix} i\left(F_D-\bar{F}_D\right) \\ F_D+\bar{F}_D \\ \sqrt{2}D_D\end{pmatrix}\end{equation}
where $D$ and $D_D$ are the usual $\theta\tilde{\theta}$ coefficients of ${\cal{A}}$ and ${\cal{A}}_D$, respectively.
The action for the theory \eqref{modspaceact} with superpotential \eqref{FIsup} can now be written as
\begin{equation}S\sim \frac{1}{2}\im\left[\int\,d^4x\,d^2\theta\,d^2\tilde{\theta}\left({\cal{F}}({\cal{A}}_i)-{\cal{A}}^i{\cal{A}}_{D\,i}\right)\right]+\frac{1}{2}\re \int\,d^4x\,\left(E_iY^i+M^iY_{D\,i}\right)\label{N2act}\end{equation}
with{\footnote{Note that we only consider electric and magnetic $F$ terms here, setting the coefficients of all $D$ terms to zero.  This choice explicitly breaks the $SU(2)_R$ invariance of the theory.}}
\begin{equation}\label{EYvectors}E_i = \begin{pmatrix} \im e_i \\ \re e_i \\ 0\end{pmatrix}\qquad M^i=\begin{pmatrix}\im m^i \\ \re m^i \\ 0\end{pmatrix}\end{equation}
To recover the ${\cal{N}}=1$ version of the action, we note that integrating out ${\cal{A}}_D$ imposes the reducing constraint on ${\cal{A}}$, up to a subtlety involving $m^j$ that we will address later.

The first term of \eqref{N2act} is manifestly invariant under the full ${\cal{N}}=2$ supersymmetry while the terms involving $Y$ and $Y_D$ also look invariant because we are used to $F$- and $D$-terms transforming into total derivatives under supersymmetry transformations.  However, as explained in Appendix A, the fact that ${\cal{A}}$ is not a reduced ${\cal{N}}=2$ superfield means that the supersymmetry transformations of $Y$ instead involve some of the extra auxiliary fields.  More specifically, the presence of $e_iF^i$ in the Lagrangian breaks the supercharge $\tilde{Q}_{\alpha}$ associated to the $\tilde{\theta}$ coordinates while the presence of $\bar{e}_i\tilde{F}^i$ breaks the supercharge $Q_{\alpha}$ associated to the $\theta$ coordinates.

Because the first term of \eqref{N2act} has a piece that is linear in both ${\cal{A}}$ and ${\cal{A}}_D$, we can try to remove this breaking by absorbing the $F^j$ and $\tilde{F}^j$ parts of $Y^j$ through an appropriate shift of ${\cal{A}}_D$.  Written in component form, the relevant part of the action is
\begin{eqnarray}S&=&
\ldots -\frac{1}{2}\im\int d^4x\,d^2\theta\,d^2\tilde{\theta}\,{\cal{A}}^i
{\cal{A}}_{D\,i}+\frac{1}{2}\re\int d^4x \left(E_iY^i +M^iY_{D\,i}\right)\nonumber \\
&=&\ldots +\frac{1}{2}\re\int d^4x\,\left[iF_{D\,j}\left(\tilde{F}^j-\bar{F}^j\right)+\bar{e}_j\left(\tilde{F}^j+\bar{F}^j\right)+2m^jF_{D\,j}\right]\label{FDterms}
\end{eqnarray}
From this, we see that it is possible to absorb the $\tilde{F}^j$ terms, and hence restore invariance with respect to $Q_{\alpha}$, by shifting $F_{D\,j}\rightarrow F_{D\,j}+i\bar{e}_j$.  Alternatively, we can absorb the $F^j$ terms, restoring invariance with respect to $\tilde{Q}_{\alpha}$, by shifting $F_{D\,j}\rightarrow F_{D\,j}-i\bar{e}_j$.  Note that it is impossible to simultaneously absorb both sets of terms, so we cannot realize simultaneous invariance with respect to both $Q_{\alpha}$ and $\tilde{Q}_{\alpha}$ in a standard manner{\footnote{The inability to do so can ultimately be traced to the reducing constraint for ${\cal{A}}_{D\,j}$ which, among other things, is responsible for the fact that its $\theta^2$ and $\tilde{\theta}^2$ components, $F_{D\,j}$ and $\bar{F}_{D\,j}$ respectively, are indeed complex conjugates of one another.}}.

Nevertheless, as discussed in more detail in Appendix A, it is possible to realize both supercharges if we modify the action of one of them to include possible inhomogeneous terms in the transformation laws of fields.  To see this, let us adopt for clarity the standard shift $F_{D\,j}\rightarrow F_{D\,j}+i\bar{e}_j$, which restores $Q_{\alpha}$-invariance and effectively sets
\begin{equation}\re\int\,d^4x\,E_iY^i \rightarrow \re\int\,d^4x\, e_jF^j\end{equation}
Under $\tilde{Q}_{\alpha}$, this term transforms into the $\tilde{\theta}\theta^2$ auxiliary component of the unreduced superfield ${\cal{A}}$.  We can cancel this using the transformation of ${\cal{A}}^i{\cal{A}}_{D\,i}$, though, if we add an inhomogeneous term proportional to $\tilde{\theta}$ to the action of $\tilde{Q}_{\dot{\alpha}}$ on ${\cal{A}}_D$ \cite{Ivanov:1997mt}.  In this manner, we are able to demonstrate that the action \eqref{N2act} is in fact invariant under a full ${\cal{N}}=2$ supersymmetry, though we can linearly realize at most half of it{\footnote{Similar structure also appears in the Abelian ${\cal{N}}=2$ Born-Infeld theories of \cite{KetovA}, which admit a nonlinear realization of an additional ${\cal{N}}=2$ supersymmetry.  Requiring such a nonlinear realization to exist also provides a guiding principle for constructing suitable non-Abelian extensions \cite{KetovNA}.  For a review, see \cite{Ketov:2001dq}.}}.  That such a nonlinear realization of a subset of supercharges is possible despite the general arguments of \cite{Witten:1981nf} was first pointed out in \cite{Hughes:1986dn}.

\subsection{Conditions for Supersymmetry-Preserving Vacua}\label{subsec:SUSYconds}

To study the conditions for a given vacuum to preserve some fraction of the ${\cal{N}}=2$ supersymmetry, it is sufficient to look at the transformation laws of the fermions in \eqref{N2act}.  Grouping the supercharges $Q,\tilde{Q}$ and fermions $\psi^j,\lambda^j$ into $SU(2)_R$ doublets ${\cal{Q}}_I=(Q,\tilde{Q})^T$ and $\Psi^j_I=(\psi^j,\lambda^j)^T$, we can write these simply as
\begin{equation}\epsilon^{\alpha\,K}{\cal{Q}}_{\alpha\,K}\Psi^j_{\beta\,I}\sim \epsilon_{IJ}\left({\cal{M}}^{(j)}\right)^J_K\epsilon^K_{\beta}\end{equation}
where{\footnote{In general, ${\cal{M}}^{(j)}$ will take the form $\begin{pmatrix} D^j & \tilde{F}^j \\ F^j & -D^j\end{pmatrix}$ but we have set $D^j=0$ because we only consider adding electric and magnetic $F$-terms to the theory.}}
\begin{equation}\label{Mmatrix}\left({\cal{M}}^{(j)}\right)^J_K=\begin{pmatrix}0 & \tilde{F}^j \\ F^j & 0\end{pmatrix}\end{equation}
Consequently, we see that a vacuum preserves the $Q$ ($\tilde{Q}$) supercharges when the expectation values of the $F^j$ ($\tilde{F}^j$) vanish for all $j$.  To compute these expectation values, we start by integrating out ${\cal{A}}_D$.
All of the terms required for this are written in \eqref{FDterms} so it is easy to see that the result is simply to set
\begin{equation}\tilde{F}^j=\bar{F}^j+2im^j\end{equation}
The expectation value of $F^j$ is then obtained by studying the $F$-term potential
\begin{equation}\frac{1}{2}\im\left(\tilde{F}^j\tau_{jk}F^k\right)+\frac{1}{2}\re\left(\bar{e}_j(\tilde{F}^j+\bar{F}^j)\right)\label{Ftermpot}\end{equation}
and concluding that
\begin{equation}\begin{split}\bar{F}^j&=-\left(\im\tau^{-1}\right)^{jk}\left(e_k+m^{\ell}\tau_{\ell k}\right)\\
\tilde{F}^j&=-\left(\im\tau^{-1}\right)^{jk}\left(e_k+m^{\ell}\bar{\tau}_{\ell k}\right)\end{split}\label{susyconds}\end{equation}
Consequently, we see that vacua for which $(e+\tau m)_j=0$ preserve the $Q$ supercharges while vacua for which $(e+\bar{\tau}m)=0$ preserve the $\tilde{Q}$ supercharges.

\subsection{SUSY or non-SUSY at $a^i_0$}

Returning to the theory of section \ref{sec:KNC}, if we recall that the combination of electric and magnetic FI terms that arose had coefficients $e_i$ and $m^j$ satisfying $(e+\bar{\tau}_0m)_j=0$, it immediately follows that the vacuum at $a^i_0$ is a supersymmetric one which preserves the non-manifest $\tilde{Q}$ supercharges.  When the superpotential is truncated as in \cite{Ooguri:2007iu}, however, we break invariance under $\tilde{Q}$ at the level of the action and the $a^i_0$ then become supersymmetry-breaking vacua.

At first glance this might seem strange because the higher order terms of \eqref{KNCdef} that we neglect when truncating do not affect the value of the scalar potential at $\tau_0$, which is given by
\begin{equation}V = \bar{k}_i\left(\im\tau_0^{-1}\right)^{ij}k_j=\overline{\left(e_i+m^j\tau_{0\,ki}\right)}\left(\im\tau_0^{-1}\right)^{ij}\left(e_j+m^{\ell}\tau_{0\,\ell j}\right)\label{vacenerg}\end{equation}
Because this quantity is manifestly positive{\footnote{The combination of $e_i+m^k\bar{\tau}_{0\,ki}=0$ and $\im\tau_0>0$ imply that $e_i+m^k\tau_{0\,ki}\ne 0$.}}, our intuition suggests that $a^i_0$ should be a supersymmetry-breaking vacuum.

It is important to note, however, that having positive energy \eqref{vacenerg} is not sufficient for a vacuum to be supersymmetry-breaking because we are in principle free to shift our definition of energy by a constant amount.  It is the specific quantity that appears in the supersymmetry algebra which matters and to determine this may require a bit more work.  In the truncated theory, the situation is actually pretty simple because there are vacua at the singular points in moduli space which preserve the manifest ${\cal{N}}=1$ supersymmetry.  Setting the energy of these vacua to zero fixes any ambiguity and leaves us with the result \eqref{vacenerg}.

The theory with full superpotential \eqref{KNCsup}, on the other hand, exhibits no such vacua.  The reason for this is that the superpotential is singular at the degeneration points.  In fact, the full superpotential is actually multivalued on the moduli space with branch points where the supersymmetric vacua of the truncated theory would otherwise be.  It was already noted in \cite{Ooguri:2007iu} that the K\"ahler normal coordinates might not be globally defined and our connection to FI terms makes this explicit.  This change in the global structure of the theory suggests that we have to re-examine our definition of energy.  To do so, let us start with the ${\cal{N}}=2$ formulation \eqref{N2act}.  In the conventional approach, where the $Q_{\alpha}$ supercharges are linearly realized, we shift $F_{D\,j}\rightarrow F_{D\,j}+i\bar{e}_j$ in the action \eqref{N2act} which effectively removes the $\tilde{F}^j$ from the second term of \eqref{Ftermpot}.  In this case, the scalar potential is easily seen to be
\begin{equation}V = \overline{\left(e_i+m^j\tau_{ki}\right)}\left(\im\tau^{-1}\right)^{ij}\left(e_j+m^{\ell}\tau_{\ell j}\right)\label{oldenerg}\end{equation}
in accordance with our result for the energy \eqref{vacenerg} of the $a^i_0$ vacuum above.  That this quantity fails to vanish at $a^i_0$ simply means that the ${\cal{N}}=1$ supersymmetry generated by the $Q_{\alpha}$ is broken there.

On the other hand, to linearly realize the $\tilde{Q}_{\alpha}$ supercharges, we saw before that it is necessary to instead shift $F_{D\,j}\rightarrow F_{D\,j}-i\bar{e}_j$ in \eqref{N2act}.  This effectively removes the $\bar{F}^j$ from the second term of \eqref{Ftermpot}, leading to the scalar potential
\begin{equation}\begin{split}\tilde{V}&=\overline{\left(e_i+m^k\bar{\tau}_{ki}\right)}\left(\im\tau^{-1}\right)^{ij}\left(e_j+m^{\ell}\bar{\tau}_{\ell j}\right)\\
&=V+4\im\left(\bar{e}_im^i\right)\label{altenerg}\end{split}\end{equation}
In other words, if we choose to linearly realize the ${\cal{N}}=1$ supersymmetry preserved by the vacuum at $a^i_0$, the definition of energy \eqref{altenerg} appropriate for that choice differs from \eqref{oldenerg} by a constant shift{\footnote{From the analysis of section \ref{subsec:SUSYconds}, we also see that it is the vanishing of $\tilde{V}$ that is required for preservation the corresponding ${\cal{N}}=1$ supersymmetry.}}.  As expected, this suitably-defined energy vanishes at $a^i_0$.

\subsection{Inclusion of $D$-terms}
To this point, we have only considered the addition of $F$-terms to the theory.  The motivation for such a restriction is that the superpotential \eqref{KNCsup} constructed from K\"ahler normal coordinates generates only these.  Nevertheless, one can also consider the addition of $D$ terms to the story.  We digress for a moment to consider this case and argue that generically all supersymmetry is broken.   This situation has recently been discussed by \cite{Aganagic:2007zm} in the context of IIB constructions and is included here only for completeness.

In general, FI-terms are characterized by the $2(N-1)$ 3-vectors $\vec{E}_j$ and $\vec{Y}^j$ of \eqref{EYvectors} which transform under $SU(2)_R$.  With our choice of basis, non-vanishing $D$-terms correspond to $\vec{E}_j$ and/or $\vec{Y}_j$ having nonzero third components{\footnote{In general, we will have nonzero $D$-terms for all choices of basis if $\vec{E}_i$ and $\vec{Y}^j$ are not all coplanar.}}.  Let us now suppose, for a moment, that a supersymmetric vacuum exists.  Using an $SU(2)_R$ rotation, we can change the original supercharges $Q_1$ and $Q_2$ into another set of supercharges $Q'_1$ and $Q'_2$ such that the $Q'_1$ annihilate the vacuum. Generalizing \eqref{Mmatrix} along with \eqref{susyconds}, the transformation matrix ${\cal{M}}$ is now given by
\begin{equation}\left({\cal{M}}^{(j)}\right)^J_K=-\left(\im\tau^{-1}\right)^{jk} \begin{pmatrix} \xi_k+\xi_D^l \bar{\tau}_{lk} & e_k+m^l \bar{\tau}_{lk} \\ \bar{e}_k+\bar{m}^l \bar{\tau}_{lk} & -\xi_k- \xi_D^l \bar{\tau}_{lk} \end{pmatrix}\;.\end{equation}
where $\xi_k$ and $\xi_D^l$ are real and generically nonzero.  Since we assume that $Q'_1$ annihilates the vacuum, the vector $\begin{pmatrix}1 \\ 0\end{pmatrix}$ should be annihilated by $\left({\cal{M}}^{(j)}\right)^J_K$ for all $j$. To that end, we want
\begin{equation}
\xi_k+\xi_D^l \bar{\tau}_{lk}=0 \qquad \text{and} \qquad e_k+m^l \tau_{lk}=0 \qquad \text{for all $k$}\;.
\end{equation}
The first condition cannot be satisfied, though, unless $\xi_k=\xi_D^k=0$ for all $k$ since $\im \tau_{lk}$ is positive definite. Therefore, in $\mathcal{N}=2$ supersymmetric language, a necessary condition to have a supersymmetric vacuum is that the $2(N-1)$ vectors $\vec E_j$ and $\vec M^j$ lie on a common plane. For $SU(2)$ gauge theory, this is always possible since there are only two vectors, but for higher gauge groups, generic FI terms necessarily break all of the supersymmetry.

\section{Critical Points, Stability, and Non-supersymmetric Vacua}

It should now be clear that the theory obtained by adding a superpotential \eqref{KNCsup} constructed from exact K\"ahler normal coordinates is significantly different from that obtained by truncating the series \eqref{KNCdef}.
This also suggests that the vacuum structure away from $a^i_0$ may be fundamentally different as well.

This opens up a new problem, though, namely to understand the full vacuum structure of theories of the form \eqref{modspaceact} in the presence of superpotentials
\begin{equation}W=e_ia^i+m^ia_{D\,i}\label{genFIsup}\end{equation}
for generic choices of $e_i$ and $m^j$.  In this section, we will take some initial steps along these lines.  More specifically, we classify non-supersymmetric critical points, study the conditions for stabilizing them, and demonstrate that, in the simple example of a rank two gauge group, one can engineer stable vacua which break the full ${\cal{N}}=2$ supersymmetry in part of the perturbative regime by choosing the $e_i$ and $m^j$ appropriately.

\subsection{Stability Conditions and Supersymmetric Vacua}

The principal object that controls the vacuum structure is the scalar potential constructed from \eqref{genFIsup}.  To start, let us write it in a covariant manner with respect to the K\"ahler metric $g_{i\bar{j}}$ of the Coulomb branch
\begin{equation}V = \left(\overline{\nabla}_{\bar{i}}\overline{W}\right)g^{\bar{i}j}\left(\nabla_j W\right)\end{equation}
Critical points of this potential satisfy
\begin{equation}\nabla_k V = \left(\overline{\nabla}_{\bar{i}}\overline{W}\right)g^{\bar{i}j}\left(\nabla_k\nabla_j W\right)=0\label{critcond}\end{equation}
while stability is determined by studying the second partials
\begin{equation}\begin{split}
\overline{\nabla}_{\bar{\ell}}\nabla_k V &= \left(\overline{\nabla}_{\bar{\ell}}\overline{\nabla}_{\bar{i}}\overline{W}\right)g^{\bar{i}j}\left(\nabla_k\nabla_j W\right) + \left(\overline{\nabla}_{\bar{i}}\overline{W}\right)g^{\bar{i}j} R^m_{\,\,\,j\bar{\ell}k}\left(\nabla_m W\right)\\
\nabla_{\ell}\nabla_k V &= \left(\overline{\nabla}_{\bar{i}}\overline{W}\right)g^{\bar{i}j}\left(\nabla_{\ell}\nabla_k\nabla_j W\right).\label{stabconds}
\end{split}\end{equation}
From this, we see that the easiest way to find critical points is to impose either $\nabla_iW=0$ or $\nabla_k\nabla_jW=0$.  For the former, it immediately follows from \eqref{stabconds} that the resulting critical points are stable.  For the latter, the same is also true at generic points provided $\nabla_{\ell}\nabla_k\nabla_j W$ also vanishes because the $R^m_{\,\,\,j\bar{\ell}k}$ term of \eqref{stabconds} is positive (semi-)definite.

These two types of vacua are in fact nothing other than the supersymmetric ones we have studied thus far.  To see this, we simply evaluate $\nabla_iW$ and $\nabla_j\nabla_k W$ in special coordinates, for which $g_{i\bar{j}}=\im\tau_{ij}$.  Because the only nonvanishing Christoffel connections are
\begin{equation}\Gamma^i_{jk}=-g_{k\bar{\ell}}\partial_jg^{i\bar{\ell}}\end{equation}
and their conjugates, this is particularly simple and results in
\begin{equation}\begin{split}\nabla_i W &= e_i+\tau_{ij}m^j\\
\nabla_i\nabla_j W &= -\frac{1}{2i} {\cal{F}}_{ijn}\left(\im\tau\right)^{nk}\left(e_k+\bar{\tau}_{ks}m^s\right)
\end{split}\end{equation}
The supersymmetric vacuum that preserves $Q$ is simply the $\nabla_i W=0$ case while the supersymmetric vacuum that preserves $\tilde{Q}$ corresponds to $\nabla_i\nabla_j W=0$.  Note that there is no issue with stability of the latter because $\nabla_k\nabla_i\nabla_j W=0$ when $(e+\bar{\tau}m)_j=0\,${\footnote{Even though this second order analysis only guarantees stability when the positive semi-definite term involving $R^m_{\,\,\,j\bar{\ell}k}$ does not have any flat directions, we know from the fact that the $(e+\bar{\tau}m)_j=0$ vacua preserve an ${\cal{N}}=1$ supersymmetry that the higher order analysis required when this condition fails must lead to stability.}}.

\subsection{Non-supersymmetric Vacua}

While it is comforting to see the supersymmetric vacua and their stability emerging naturally from this framework, it is at the same time disappointing that the simplest ways to realize critical points of the potential fail to yield anything new.

In principle, the mechanism by which new critical points of the potential can be found is quite simple.  We need a mixture of sorts of the supersymmetric and hidden supersymmetric cases where $g^{\bar{i}j}\overline{\nabla}_{\bar{i}}\overline{W}$ and $\nabla_k\nabla_j W$ are both nonzero but, in a suitable basis, have complementary components vanishing so that the contraction in \eqref{critcond} is zero.  Unfortunately, there is no apparent reduction in complexity of \eqref{stabconds} in this case so it is difficult to spell out simple conditions for a vacuum constructed in such a manner to be stable.

To describe the idea more precisely, let us drop the covariant notation of \eqref{critcond} and \eqref{stabconds} and instead re-express the various derivatives of $V$ in special coordinates as
\begin{equation}\partial_q V = -\frac{1}{2i}F^f{\cal{F}}_{qfe}\tilde{F}^e\label{critcondd}\end{equation}
and
\begin{equation}\begin{split}
\partial_p\partial_q V &= -\frac{1}{2i}F^f\left({\cal{F}}_{pqfe}-\frac{1}{2i}\left[{\cal{F}}_{pfm}\left(\im\tau^{-1}\right)^{mn}{\cal{F}}_{qne}+\left(p\leftrightarrow q\right)\right]\right)\tilde{F}^e\\
\bar{\partial}_{\bar{p}}\partial_q V &= \frac{1}{4}\left[ \bar{\tilde{F}}^a\overline{\cal{F}}_{pam}\left(\im\tau^{-1}\right)^{mn}{\cal{F}}_{qnb}\tilde{F}^b+F^a{\cal{F}}_{qam}\left(\im\tau^{-1}\right)^{mn}\overline{\cal{F}}_{pnb}\bar{F}^b\right]
\label{stabcondd}\end{split}\end{equation}
where $F^a$ and $\tilde{F}^b$ are the auxiliary field expectation values of \eqref{susyconds}.

In general, rather than looking for stable vacua at fixed $e_i$ and $m^j$, we will find it easier to reverse our thinking and approach the problem in a manner analogous to \cite{Ooguri:2007iu}.  So, we instead specify a point $u_0$ along the Coulomb branch at which we would like to engineer a stable critical point and develop an algorithm for obtaining values $e_i$ and $m^j$ that do the job, if such exist.

To aid in this task, let's first use \eqref{critcondd} and \eqref{stabcondd} to study the general structure of supersymmetry-breaking vacua.  The first thing to note is that the vectors $F^a$ and $\tilde{F}^a$ at such a vacuum can never be parallel.  The reason for this is that a critical point for which they are parallel satisfies $\bar{F}^{\bar{p}}\bar{\partial}_{\bar{p}}\partial_qVF^q=0$ from \eqref{critcondd} and \eqref{stabcondd} and
\begin{equation}
\begin{pmatrix} e^{i\phi} F^p & e^{-i\phi} \bar{F}^{\bar{p}}\end{pmatrix}
\begin{pmatrix} \partial_p \bar{\partial}_{\bar{q}} V & \partial_p \partial_{q} V \\ \bar{\partial}_{\bar{p}} \bar{\partial}_{\bar{q}} V & \partial_q \bar{\partial}_{\bar{p}} V \end{pmatrix}
\begin{pmatrix} e^{-i\phi} \bar{F}^{\bar{q}} \\ e^{i\phi} F^q\end{pmatrix}
=2\re (e^{2i\phi} F^p \partial_p \partial_q V F^q)\;,
\end{equation}
where $\phi$ is a real phase. There is always a choice of $\phi$ for which this is negative so we see that such a critical point can never be stable.  Incidentally, this means that for real $e_i$ and $m^j$, achieving a metastable supersymmetry-breaking
vacuum is impossible since $F^a=\tilde{F}^a$ in this case. Since neither $e_i+\tau_{ij} m^{j}=0$ nor $e_i+\bar{\tau}_{ij} m^{j}=0$ is not attainable either, the only possible minimum occurs when the metric is singular. That is, when we have a dyon condensation point and the dyon charge is proportional to $(e_i, m^j)$, the effective potential vanishes at that point and we have a supersymmetric vacuum there.

Let us now consider a coordinate transformation matrix $Q^i_{\,\,i'}$ under which ${\cal{F}}_{qfe}$ transforms as
\begin{equation}{\cal{F}}_{qfe}\rightarrow {\cal{F}}'_{q'f'e'}={\cal{F}}_{qfe}Q^q_{\,\,q'}Q^f_{\,\,f'}Q^e_{\,\,e'}\label{Ftrans}\end{equation}
Because $F^a$ and $\tilde{F}^a$ are not parallel, we can always perform a coordinate transformation $Q^i_{\,\,i'}$ so that the only non-vanishing component of $F^a$ ($\tilde{F}^a$) is the first (second) one.  In this basis, ${\cal{F}}_{q12}=0$ for all $q$.

It should now clear how to engineer a critical point at $u_0$ that can potentially be stabilized.  Given ${\cal{F}}$, we use a coordinate transformation \eqref{Ftrans} so that ${\cal{F}}_{q12}=0$ for all $q$.  Such a coordinate transformation should generically exist because we have $(N-1)^2$ degrees of freedom in $Q$ to satisfy only $N-1$ conditions.  With such a $Q$, we then choose values of $F^a$ and $\tilde{F}^a$ as
\begin{equation}F=Q\begin{pmatrix}\zeta \\ 0 \\ 0 \\ \vdots\end{pmatrix}\qquad\qquad \tilde{F}=Q\begin{pmatrix}0 \\ \xi \\ 0 \\ \vdots\end{pmatrix}\label{PEE}\end{equation}
Once such a choice is made, we can generically solve \eqref{PEE} for the corresponding values of $e_i$ and $m^j$ because this is a system of $2(N-1)$ linear equations in $2(N-1)$ variables.

Now that we have constructed critical points, we must turn to the question of their stability.  Given that we can actually engineer families of critical points parametrized by $\zeta$ and $\xi$, one might hope that there is enough freedom left over to achieve stability as well.  Studying this issue is very complicated in practice, though, so to demonstrate the principle in action we focus on the most basic example we can find.  It is clear that vacua of this sort cannot be generated when the gauge group has rank 1, so we turn instead to the rank 2 case of $SU(3)$ Seiberg-Witten theory.

\subsection{An $SU(3)$ Example}

In what follows, we shall work exclusively in the perturbative regime $a_i\gg \Lambda$, where the Seiberg-Witten prepotential ${\cal{F}}$ appearing in \eqref{modspaceact} takes the approximate form \cite{Argyres}
\begin{equation}{\cal{F}}(a_i)=\frac{i}{4\pi}\sum_{i<j}^3\left(a_i-a_j\right)^2\ln\left[\frac{\left(a_i-a_j\right)^2}{\Lambda^2}\right]\end{equation}
We will henceforth set $\Lambda=1$ and use the coordinate basis
\begin{equation}x=a_2-a_1\qquad y=a_3-a_2\end{equation}
In terms of these, the prepotential is given by a simple expression
\begin{equation}{\cal{F}}=\frac{i}{4\pi}\left(x^2\ln x^2+y^2\ln y^2+(x+y)^2\ln(x+y)^2\right)\end{equation}
The various derivatives we shall need when studying \eqref{critcondd} and \eqref{stabcondd} are now easily evaluated.  We start with the period matrix
\begin{equation}\tau=\frac{i}{2\pi}\begin{pmatrix}6+\ln x^2+\ln (x+y)^2 & 3+\ln(x+y)^2 \\
3+\ln(x+y)^2 & 6+\ln y^2+\ln(x+y)^2\end{pmatrix}\end{equation}
and proceed to its derivatives
\begin{equation}\partial_x\tau_{ij} = {\cal{F}}_{xij}=\frac{i}{\pi(x+y)}\begin{pmatrix}2+\frac{y}{x} & 1 \\ 1 & 1\end{pmatrix}\qquad \partial_y\tau_{ij} = {\cal{F}}_{yij}=\frac{i}{\pi(x+y)}\begin{pmatrix}1 & 1 \\ 1 & 2+\frac{x}{y}\end{pmatrix}\end{equation}
and second derivatives
\begin{equation}\begin{split}{\cal{F}}_{xxij}&=\frac{1}{i\pi (x+y)^2}\begin{pmatrix}2+2\left(\frac{y}{x}\right)+\left(\frac{y}{x}\right)^2 & 1 \\ 1 & 1\end{pmatrix}\\
{\cal{F}}_{xyij}&=\frac{1}{i\pi(x+y)^2}\begin{pmatrix}1&1\\1&1\end{pmatrix}\\
{\cal{F}}_{yyij}&=\frac{1}{i\pi (x+y)^2}\begin{pmatrix}1 & 1 \\ 1 & 2+2\left(\frac{x}{y}\right)+\left(\frac{x}{y}\right)^2\end{pmatrix}
\end{split}\end{equation}

In this simple example, we can set ${\cal{F}}_{q12}=0$ using a transformation of the form \eqref{Ftrans} with $Q$ given by
\begin{equation}Q=\begin{pmatrix}-x & x+y+\sqrt{(x+y)^2-xy} \\ x+y+\sqrt{(x+y)^2-xy} & -y\end{pmatrix}\label{Qres}\end{equation}
To find a choice of $e_i,m^j$ for which the potential has a critical point at $(x_0,y_0)$, we turn then to the equations
\begin{equation}
F=Q\begin{pmatrix}\zeta \\ 0\end{pmatrix}\qquad \tilde{F}=Q\begin{pmatrix}0 \\ \xi\end{pmatrix}\label{QEEtildeqns}
\end{equation}
where here $\zeta$ and $\xi$ are nonzero constants that we are free to choose, $Q$ is as in \eqref{Qres}, and $F,\tilde{F}$ are given in terms of $e_i,m^j,\tau_{k\ell}$ as in \eqref{susyconds}.  Given our result \eqref{Qres} for $Q$, \eqref{QEEtildeqns} is equivalent to the requirement
\begin{equation}
F=-\zeta \begin{pmatrix}-x \\ x+y+\sqrt{(x+y)^2-xy}\end{pmatrix},\quad
\tilde{F} = -\xi  \begin{pmatrix}x+y+\sqrt{(x+y)^2-xy} \\ -y\end{pmatrix}\label{EEtildsoln}
\end{equation}

As mentioned before, we generically expect that it is possible to choose $e_i,m^j$ for any nonzero choice of $\zeta$ and $\xi$ such that \eqref{EEtildsoln} is satisfied at a fixed point $(x_0,y_0)$.  From this point onward, we will assume that the situation is indeed generic and take the existence of such a solution for granted.

\subsubsection{Stability}

Engineering a critical point is one matter but achieving stability is the real challenge.  However, as we will now demonstrate through a simple scaling argument, it is possible to take advantage of the freedom to adjust $\zeta$ and $\xi$ to choose FI terms that engineer \emph{stable} supersymmetry-breaking vacua in part of the perturbative regime.

In particular, let us consider the regime $y\gg x\gg 1$.  We will now show that if we choose $\zeta$ and $\xi$ to be of order 1, the critical point constructed by solving \eqref{QEEtildeqns} is always locally stable.  Expanding the Hessian
\begin{equation}
H=\begin{pmatrix} \partial_p \bar{\partial}_{\bar{q}} V & \partial_p \partial_{q} V \\ \bar{\partial}_{\bar{p}} \bar{\partial}_{\bar{q}} V & \partial_q \bar{\partial}_{\bar{p}} V \end{pmatrix}
\end{equation}
at the critical point, it is straightforward to check whether the eigenvalues $\lambda_{1}\cdots\lambda_{4}$ of $H$ are all positive. In the limit mentioned above, $H$ scales near infinite $y$ as follows:
\begin{equation}
H=\begin{pmatrix} h_{11} y^2 & h_{12} y 	& h_{13}y    & h_{14}\\
		  h_{21} y   & h_{22}   	& h_{14}     & \frac{h_{24}}{y}\\
		  h_{31}y    & h_{41}   	& h_{11} y^2 & h_{21} y\\
		  h_{41}     & \frac{h_{42}}{y} & h_{12} y   & h_{22} \end{pmatrix}
\end{equation}
where the $h_{ij}$ depend logarithmically on $y$ (and on $x$, $\xi$, $\zeta$). To leading order in $y$, the four eigenvalues are
\begin{equation}
h_{11} y^2 \quad \text{and}\quad \frac{h_{11}h_{22}-h_{12}h_{21}}{h_{11}}\;,
\end{equation}
with multiplicity two for each. Since the matrix $\partial_p \bar{\partial}_{\bar{q}} V$ is positive definite from \eqref{stabconds} , the critical point is locally stable.

To illustrate potential subtleties that can arise when studying stability, let us also consider a second regime, namely $x\sim y\gg 1$.  If we use $r$ to denote the scale of $x$ and $y$, the quantities appearing in \eqref{stabcondd} behave at large $r$ as
\begin{equation}\begin{split}\tau_{ij}&\sim \ln r \\
{\cal{F}}_{ijk}&\sim r^{-1} \\
{\cal{F}}_{ijk\ell}&\sim r^{-2}\\
F&\sim \zeta r\\
\tilde{F} &\sim \xi r\\
\end{split}\end{equation}
This means that $\partial_p\partial_q V\sim \zeta\xi$ while $\bar{\partial}_{\bar{p}}\partial_q V \sim \zeta^2(\ln r)^{-1}+\xi^2 (\ln r)^{-1}$ at large $r$.  If we take $\zeta$ and $\xi$ to be of order 1 in this case, then the $\partial_p\partial_qV$ terms dominate and the Hessian necessarily has at least one negative eigenvalue.

Given the above scalings, though, one might naively think that stability can be achieved by taking $\zeta$ to be very large, say $\zeta\sim r$ for example, and $\xi$ to be small, as in $\xi\sim r^{-1}$, because this ensures that the dominant contribution to the Hessian comes from $\bar{\partial}_{\bar{p}}\partial_qV$.  This looks good for stability but unfortunately $\bar{\partial}_{\bar{p}}\partial_qV$ has an obvious flat direction in this case proportional to $\tilde{F}^q$ because
\begin{equation}F^a{\cal{F}}_{qam}\tilde{F}^q=0\end{equation}
The corresponding zero eigenvalue is generically lifted by the next-leading contribution to the Hessian, which comes from the off-diagonal term $\partial_p\partial_q V$.  This means that the leading correction to this zero eigenvalue is in fact negative and our critical point is actually unstable.

\section{Connection with Flux Compactifications}

Until now, we have mainly focused on the field theory perspective of Seiberg-Witten theories deformed by electric and magnetic FI terms.  Here we will briefly discuss the geometric realization of the vacua that we have studied so far in the context of string theory compactifications in the presence of NS and RR fluxes. In a series of papers \cite{GeometricEngineering}, Seiberg-Witten theories were geometrically engineered in Type IIA and IIB string theories compactified on Calabi-Yau manifolds in a rigid limit of special geometry. For example, in type IIB, $SU(N)$ Seiberg-Witten theory was realized on a geometry constructed as a $K3$ fibration over a ${\bf P}^1$ base. Near the singular locus of $K3$ over ${\bf P}^1$, the Calabi-Yau manifold becomes
$${
z+{\Lambda^{2N} \over z}+2W_{A_{N-1}}(x_1,u_i)+2x_2^2+2x_3^2=0, \label{defADE}
}$$
where $W_{A_{N-1}}(x_k,u_i)$ corresponds to the characteristic polynomial of the Seiberg-Witten theory.

Non-vanishing NS and RR fluxes, $H=H_{RR}+\tau_{\rm st} H_{NS}$, generate a superpotential \cite{GVW} and lift the vacuum degeneracy in the Calabi-Yau manifold \cite{DouglasKachru},
\begin{eqnarray}
W_{GVW}&=&\int H \wedge \Omega =\int_{B_i} H \ \int_{A_i}\Omega -\int_{A_i} H \int_{B_i} \Omega \nonumber \\
&\equiv& e_i a^i +m^i a_{D\, i} \nonumber
\end{eqnarray}
where $(A_i, B_i)=\delta_{ij}$ comprise a symplectic basis of three-cycles. Since the integrals of the holomorphic $3$-from, $\Omega$, are naturally identified with the periods of Seiberg-Witten theory while turning on generic fluxes yields a set of complex valued $(e_i, m^j)$, we can realize the model treated in this paper by adding fluxes appropriately.

Supersymmetry-breaking in Calabi-Yau compactifications of this sort have also appeared 
in connection with brane/antibrane systems in \cite{Vafa:2000wi} and more recently in
\cite{Aganagic:2006ex,Heckman:2007wk,Douglas:2007tu,Marsano:2007fe,Heckman:2007ub,Aganagic:2007kd,Aganagic:2007zm}.  In particular, a notion of geometric transition involving gauge/gravity duality was generalized to the non-supersymmetric setting, allowing configurations of branes and antibranes to be studied using the same sort of Abelian gauge theory with FI terms considered in the present paper.  Because the vacua studied here have natural realizations on the flux side in this context, it would be interesting to follow the geometric transition in reverse and study them from this perspective.

The flux realization of the model with FI terms also gives us a clear picture of how the potential behaves near a singular point in moduli space.  At such a point, a massless dyon with charges $(n^e_i,n^m_i)$ emerges and the corresponding cycle $\gamma=n^e_i A_i +n_i^m B_i$ shrinks. When we turn on generic FI-terms, the scalar potential diverges there for the simple reason that non-zero fluxes penetrate the cycle
$${
\int_{\gamma}H=n_i^e e_i-n_i^m m_i \neq 0.
}$$
and render infinite the energy cost associated with closing it up.

\section*{Acknowledgments}

We would like to thank S.~Ferrara, S.~Kachru, S.~Ketov, M.~Nitta, M.~Shigemori, and S.~Trivedi for discussions.
This research is supported in part by DOE grant DE-FG03-92-ER40701.
J.M. is also supported by a John A McCone
postdoctoral fellowship. The research of H.O. is also supported in part
by the NSF grant OISE-0403366, by the 21st Century COE Program
at the University of Tokyo, and by the Kavli Foundation.
Y.O. is also supported in part by the JSPS Fellowship
for Research Abroad. C.P. is also supported in part by
Samsung Scholarship. H.O. thanks the Aspen Center for Physics for the hospitality at the initial stage of this work.

\bigskip

\appendix

\section{${\cal{N}}=2$ Superfields, FI Terms, and Nonlinear Realization of Supersymmetry}

In this Appendix, we make explicit our superfield conventions and discuss the nonlinear realization of ${\cal{N}}=2$ supersymmetry of the action \eqref{N2act} in greater detail.  The results of this appendix are not new but we present them
in this component language for both clarity and completeness.

\subsection{${\cal{N}}=2$ Superspace and Superfields}

We shall work in the ${\cal{N}}=2$ superspace conventions of \cite{Lykken:1996xt} with two anticommuting coordinates $\theta^{\alpha}$ and $\tilde{\theta}^{\alpha}$.  The standard realization of ${\cal{N}}=2$ supersymmetry on this space is through the operators
\begin{equation}Q_{\alpha}=\frac{\partial}{\partial\theta^{\alpha}}-i(\sigma^{\mu}\bar{\theta})_{\alpha}\partial_{\mu}\qquad \tilde{Q}_{\alpha}=\frac{\partial}{\partial\tilde{\theta}^{\alpha}}-i(\sigma^{\mu}\bar{\tilde{\theta}})_{\alpha} \partial_{\mu} \label{Qops}\end{equation}
and their conjugates.  A generic chiral ${\cal{N}}=2$ superfield, ${\cal{A}}$ can be constructed from two chiral ${\cal{N}}=1$ superfields, $\Phi$ and $G$, along with a chiral ${\cal{N}}=1$ spinor superfield $W_{\alpha}$ as
\begin{equation}{\cal{A}}(\tilde{y},\theta,\tilde{\theta})=\Phi(\tilde{y},\theta)+i\sqrt{2}\tilde{\theta}W(\tilde{y},\theta)
+\tilde{\theta}^2G(\tilde{y},\theta)\label{N2supexp}\end{equation}
where $\tilde{y}^{\mu}=x^{\mu}+i\theta\sigma^{\mu}\bar{\theta}+i\tilde{\theta}\sigma^{\mu}\bar{\tilde{\theta}}$.  The ${\cal{N}}=1$ superfields admit further component expansions
\begin{equation}\begin{split}\Phi(\tilde{y},\theta)&=\phi(\tilde{y})+\sqrt{2}\theta\psi(\tilde{y})+\theta^2F(\tilde{y})\\
W_{\alpha}(\tilde{y},\theta)&=-i\lambda_{\alpha}(\tilde{y})+\theta_{\gamma}\left(\delta^{\gamma}_{\alpha}D(\tilde{y})-\frac{i}{2}\left(\sigma^{\mu}\bar{\sigma}^{\nu}\theta\right)^{\gamma}_{\alpha}F_{\mu\nu}(\tilde{y})\right)-i\theta^2\xi_{\alpha}(\tilde{y})\\
G(\tilde{y},\theta)&= \tilde{F}(\tilde{y})+\sqrt{2}\theta\eta(\tilde{y})+\theta^2C(\tilde{y})
\end{split}\end{equation}
Note that $W_{\alpha}$ does
not satisfy any constraints so it is not quite the superfield with which we are
used to constructing ${\cal{N}}=1$-invariant actions.  In particular, $F_{\mu\nu}$ does not satisfy the Bianchi identity and $\xi_{\alpha}$ is not proportional
to $\left(\sigma^{\mu}\partial_{\mu}\bar{\lambda}\right)_{\alpha}$.

Let us now consider a chiral superfield ${\cal{A}}_D$ satisfying the additional
reducing constraint \eqref{redconst}, which we repeat here for convenience
\begin{equation}\left(D^{a\,\alpha}D^b_{\alpha}\right){\cal{A}}_D=\left(\overline{D}^a_{\dot{\alpha}}\overline{D}^{b\,\dot{\alpha}}\right){\cal{A}}_D^{\dag}\end{equation}
This again admits an expansion of the sort \eqref{N2supexp}
\begin{equation}{\cal{A}}_D(\tilde{y},\theta,\tilde{\theta})=\Phi_D(\tilde{y},\theta)+i\sqrt{2}\tilde{\theta}W_D(\tilde{y},\theta)+\tilde{\theta}^2G_D(\tilde{y},\theta)\end{equation}
The corresponding ${\cal{N}}=1$ expansion, though, becomes
\begin{equation}\begin{split}\Phi_D(\tilde{y},\theta)&=\phi_D(\tilde{y})+\sqrt{2}\theta\psi_D(\tilde{y})+\theta^2F_D(\tilde{y})\\
W_{\alpha\,D}(\tilde{y},\theta)&=-i\lambda_{\alpha\,D}(\tilde{y})+\theta_{\gamma}\left(\delta_{\alpha}^{\gamma}D_D(\tilde{y})-\frac{i}{2}\left(\sigma^{\mu}\bar{\sigma}^{\nu}\theta\right)_{\alpha}^{\gamma}F_{\mu\nu\,D}(\tilde{y})\right)+\theta^2\sigma^{\mu}_{\alpha\dot{\beta}}\partial_{\mu}\bar{\lambda}_D^{\dot{\beta}}(\tilde{y})\\
G_D(\tilde{y},\theta)&=\bar{F}_D(\tilde{y})+i\sqrt{2}\left(\theta\sigma^{\mu}\partial_{\mu}\bar{\psi}_D(\tilde{y})\right)-\theta^2\partial^2\bar{\phi}_D(\tilde{y})
\end{split}\end{equation}
with
\begin{equation}\partial_{[\mu}F_{\nu\rho]\,D}=0\end{equation}
Note that the reduced ${\cal{N}}=2$ superfield differs from the unconstrained one in several ways.  In addition to $F_{\mu\nu\,D}$ satisfying the Bianchi identity, the $\theta^2$ component of $W_{\alpha\,D}$ as well as the $\theta$ and $\theta^2$ components of $G_D$ are no longer independent but instead are given by total derivatives of other component fields.  Finally, the bottom component of $G_D$ is an auxiliary field which is not independent but instead set to the complex conjugate of the top component of $\Phi_D$, denoted $F_D$.

\subsection{Supersymmetry Transformations of the Action \eqref{N2act}}

We now turn our attention to the action \eqref{N2act}, repeated here for convenience
\begin{equation}S=\frac{1}{2}\im\left[\int\,d^4x\,d^2\theta\,d^2\tilde{\theta}\left({\cal{F}}({\cal{A}}^i)-{\cal{A}}^i{\cal{A}}_{D\,i}\right)\right]+\frac{1}{2}\re\int\,d^4x\left(E_iY^i+M^iY_{D\,i}\right)\label{N2actt}\end{equation}

With the above expansions of ${\cal{A}}$ and ${\cal{A}}_D$, it is straightforward to write \eqref{N2actt} in components and demonstrate that integrating out ${\cal{A}}_D$ when $E_i=M^j=0$ indeed simply causes ${\cal{A}}$ to become a reduced ${\cal{N}}=2$ chiral superfield.  The situation of vanishing $E_i$ and $M^j$ is also one in which the action \eqref{N2act} clearly preserves ${\cal{N}}=2$ supersymmetry because it is simply the top component of an ${\cal{N}}=2$ chiral superfield, which transforms into a total derivative.  What we will focus on in the remainder of this appendix, though, is the realization of ${\cal{N}}=2$ supersymmetry when $E_i$ and $M^j$ are nonzero.

Written in component form, the FI terms in \eqref{N2actt} are given by
\begin{equation}\frac{1}{2}\re\int\,d^4x\left(E_iY^i+M^iY_{D\,i}\right)=\frac{1}{2}\int\,d^4x\left\{\re\left(e_iF^i+\bar{e}_i\tilde{F}^i\right)+\re \left(2mF_D\right)\right\}\label{FIexpF}\end{equation}
Because $F_D$ is the $\theta^2$ component of a reduced ${\cal{N}}=2$ superfield, it is easy to see that it transforms into a total derivative under the action of all
${\cal{N}}=2$ generators \eqref{Qops}.  This is not so for $F$ and $\tilde{F}$,
though, as there are two problematic non-derivative transformations
\begin{equation}\epsilon\tilde{Q}F=\sqrt{2}\epsilon\xi\qquad \epsilon Q\tilde{F}=\sqrt{2}\epsilon\eta\label{probtrans}\end{equation}
If ${\cal{A}}$ were reduced, the $\xi_{\alpha}$ would be proportional to $(\sigma^{\mu}\partial_{\mu}\bar{\lambda})_{\alpha}$ while $\eta_{\alpha}\sim (\sigma^{\mu}\partial_{\mu}\bar{\psi})_{\alpha}$.  In this case, the RHS of \eqref{probtrans} would consist of total derivatives and invariance of \eqref{FIexpF} would be assured.  As it stands, however, \eqref{FIexpF} is not preserved by either $Q_{\alpha}$ or $\tilde{Q}_{\alpha}$.

As mentioned in the text, we can try to improve the situation by suitably adjusting the transformation law of ${\cal{A}}_D$.  In particular, because ${\cal{A}}_D$ appears in \eqref{N2act} only via a term which is linear in both ${\cal{A}}$
and ${\cal{A}}_D$, we can try to absorb the terms on the RHS of \eqref{probtrans} by suitably shifting the transformation laws of component fields of ${\cal{A}}_D$.  Indeed, expanding in components we see that $\xi^i$ and $\eta^i$ appear in the ${\cal{A}}{\cal{A}}_D$ term as
\begin{equation}\int\,d^2\theta\,d^2\tilde{\theta}\,\left(-{\cal{A}}^i{\cal{A}}_{D\,i}\right)=\ldots +\frac{1}{2}\im\left(\lambda_{D\,i}\xi^i+\psi_{D\,i}\eta^i\right)+\ldots\end{equation}
This means that the full action \eqref{N2actt} can be made invariant under the full ${\cal{N}}=2$ supersymmetry if we modify the transformation laws of $\lambda_D$ and $\psi_D$ under $\tilde{Q}$ and $Q$ from
\begin{equation}\left(\epsilon\tilde{Q}\right)\lambda_{\alpha\,D}=\sqrt{2}\epsilon_{\alpha}\bar{F}_D\qquad\left(\epsilon Q\right)\psi_{\alpha\,D}=\sqrt{2}\epsilon_{\alpha}F_D\end{equation}
to
\begin{equation}\left(\epsilon\tilde{Q}\right)\lambda_{\alpha\,D}=\sqrt{2}\epsilon_{\alpha}\left(\bar{F}_D-ie\right)\qquad \left(\epsilon Q\right)\psi_{\alpha\,D}=\sqrt{2}\left(F_D-i\bar{e}\right)\end{equation}
Because of the inhomogeneous terms now present in these transformation laws, the
realization is no longer linear.  We can linearize one of the supercharges, though, by shifting $F_D$ appropriately.  In particular, if we take $F_D\rightarrow
F_D+i\bar{e}$ then the $Q$ transformation of $\psi_{\alpha\,D}$ becomes linear.  As shown in the text, this can be understood by noting that such a shift also effectively removes the $e_iF^i$ term from the action.
 On the other hand, taking $F_D\rightarrow F_D-i\bar{e}$ renders the $\tilde{Q}$ transformation of $\lambda_{\alpha\,D}$ linear.  In this case, the shift of $F_D$ effectively removes the $\bar{e}_i\tilde{F}^i$ term from the action.  Note that, while we have a choice to linearly realize either supersymmetry, it is impossible to simultaneously do so for the full ${\cal{N}}=2$ supersymmetry algebra.  This implies that at most ${\cal{N}}=1$ supersymmetry can be realized in vacua of the model \eqref{N2actt}.

\end{document}